\documentstyle[12pt,epsf]{article}
\begin{document}
\setlength{\headheight}{0in}
\setlength{\headsep}{0in}
\setlength{\topskip}{1ex}
\setlength{\textheight}{8.5in}
\setlength{\topmargin}{0.5cm}
\setlength{\baselineskip}{0.24in}
\catcode`@=11
\long\def\@caption#1[#2]#3{\par\addcontentsline{\csname
  ext@#1\endcsname}{#1}{\protect\numberline{\csname
  the#1\endcsname}{\ignorespaces #2}}\begingroup
    \small
    \@parboxrestore
    \@makecaption{\csname fnum@#1\endcsname}{\ignorespaces #3}\par
  \endgroup}
\catcode`@=12
\def\slashchar#1{\setbox0=\hbox{$#1$}           
   \dimen0=\wd0                                 
   \setbox1=\hbox{/} \dimen1=\wd1               
   \ifdim\dimen0>\dimen1                        
      \rlap{\hbox to \dimen0{\hfil/\hfil}}      
      #1                                        
   \else                                        
      \rlap{\hbox to \dimen1{\hfil$#1$\hfil}}   
      /                                         
   \fi}                                         %
\newcommand{\newc}{\newcommand}
\newc{\gsim}{\lower.7ex\hbox{$\;\stackrel{\textstyle>}{\sim}\;$}}
\newc{\lsim}{\lower.7ex\hbox{$\;\stackrel{\textstyle<}{\sim}\;$}}
\newc{\mtpole}{M_t}
\newc{\mbpole}{M_b}
\newc{\mtop}{m_t}
\newc{\mbot}{m_b}
\newc{\mz}{m_Z}
\newc{\mw}{m_W}
\newc{\alphasmz}{\alpha_s(m_Z^2)}
\newc{\MS}{\hbox{\rm\overline{MS}}}  \newc{\msbar}{\MS}
\newc{\DR}{\hbox{\rm\overline{DR}}}
\newc{\tbeta}{\tan\beta}
\newc{\stopL}{{\widetilde t_L}}
\newc{\stopR}{{\widetilde t_R}}
\newc{\stopone}{\widetilde t_1}
\newc{\stoptwo}{\widetilde t_2}

\newc{\BR}{\hbox{\rm BR}}
\newc{\xpb}{\hbox{\rm\, pb}}
\newc{\zbb}{Z\to b\bar}
\newc{\Gb}{\Gamma (Z\to b\bar b)}
\newc{\Gh}{\Gamma (Z\to \hbox{\rm had})}
\newc{\rbsm}{R_b^\hbox{\rm sm}}
\newc{\rbsusy}{R_b^\hbox{\rm susy}}
\newc{\drb}{\delta R_b}
%
\newc{\xc}{\chi^{\pm}}
\newc{\xn}{\chi^{0}}
\newc{\swsq}{\sin^2\theta_W}
\newc{\tw}{\tan\theta_W}
\newc{\cw}{\cos\theta_W}
\newc{\sw}{\sin\theta_W}
\newc{\mhp}{m_{H^\pm}}
\newc{\mhalf}{m_{1/2}}
\newc{\ie}{{\it i.e.}}		
\newc{\etal}{{\it et al.}}
\newc{\eg}{{\it e.g.}}		
\newc{\ev}{\hbox{\rm\,eV}}		
\newc{\kev}{\hbox{\rm\,keV}}		
\newc{\mev}{\hbox{\rm\,MeV}}		
\newc{\gev}{\hbox{\rm\,GeV}}		
\newc{\tev}{\hbox{\rm\,TeV}}
\newc{\hbeta}{{N_1}}
\def\order#1{{\cal O}(#1)}
\def\mass#1{m_{#1}}
\def\Mass#1{M_{#1}}
\def\NPB#1#2#3{Nucl. Phys. {\bf B#1} (19#2) #3}
\def\PLB#1#2#3{Phys. Lett. {\bf B#1} (19#2) #3}
\def\PLBold#1#2#3{Phys. Lett. {\bf#1B} (19#2) #3}
\def\PRD#1#2#3{Phys. Rev. {\bf D#1} (19#2) #3}
\def\PRL#1#2#3{Phys. Rev. Lett. {\bf#1} (19#2) #3}
\def\PRT#1#2#3{Phys. Rep. {\bf#1} (19#2) #3}
\def\ARAA#1#2#3{Ann. Rev. Astron. Astrophys. {\bf#1} (19#2) #3}
\def\ARNP#1#2#3{Ann. Rev. Nucl. Part. Sci. {\bf#1} (19#2) #3}
\def\MODA#1#2#3{Mod. Phys. Lett. {\bf A#1} (19#2) #3}
\def\ZPC#1#2#3{Zeit. f\"ur Physik {\bf C#1} (19#2) #3}
\def\APJ#1#2#3{Ap. J. {\bf#1} (19#2) #3}
\def\MPL#1#2#3{Mod. Phys. Lett. {\bf A#1} (19#2) #3}
\def\beq{\begin{equation}}
\def\eeq{\end{equation}}
\def\bea{\begin{eqnarray*}}
\def\eea{\end{eqnarray*}}

%
\begin{titlepage}
\begin{flushright}
{\setlength{\baselineskip}{0.18in}
{\normalsize
hep-ph/9603336 \\
UM-TH-96-04\\
SLAC-PUB-7131\\
March 1996\\
}}
\end{flushright}
\vskip 2cm
\begin{center}

{\Large\bf 
 Higgsino Cold Dark Matter Motivated by Collider Data }

\vskip 1cm

{\large
G.L.~Kane${}^a$ 
and James D.~Wells${}^{b,}$\footnote{Work supported by 
the Department of Energy, contract DE-AC03-76SF00515.} \\}

\vskip 0.5cm
{\setlength{\baselineskip}{0.18in}
{\normalsize\it ${}^a$Randall Physics Laboratory \\
           University of Michigan \\
	   Ann Arbor, MI 48109--1120 \\}
\vskip 4pt
{\normalsize\it ${}^b$Stanford Linear Accelerator Center  \\
     Stanford University, Stanford, CA 94309 \\} }

\end{center}
\vskip .5cm
\begin{abstract}

Motivated by the supersymmetric interpretation of the CDF 
$ee\gamma\gamma + \slashchar{E}_T$ event
and the reported $Z\to b\bar b$ excess at LEP, we analyze the
Higgsino as a cold dark matter candidate.  We examine the 
constraints as implied by the collider experiments,
and then calculate its relic density.  We find that this Higgsino-like
lightest supersymmetric particle is a viable cold dark matter candidate
($0.05< \Omega h^2 < 1$), and we discuss its favorable 
prospects for laboratory detection.

\end{abstract}
\end{titlepage}

\setcounter{footnote}{0}
\setcounter{page}{1}
\setcounter{section}{0}
\setcounter{subsection}{0}
\setcounter{subsubsection}{0}


\section*{Introduction}

One of the successes of the supersymmetric version of the standard model
is that it provides a natural candidate for the cold dark matter of the
universe.  Many authors~\cite{diehl95} have lauded this feature of
supersymmetry.  Studies often assume that
the lightest supersymmetric particle (LSP) is the bino ($\tilde B$), the
supersymmetric partner of the hypercharge gauge boson, for two reasons.
Bino annihilation can give about the right relic density.  Second,
minimal supergravity models with gauge coupling unification and with
universal scalar and gaugino masses at the unification scale generically
produce a Bino LSP after the full renormalization group flow of all
sparticle masses, Yukawas, and gauge couplings~\cite{CMSSM}.  In
general, Higgsino-like LSP's annihilate too efficiently to provide
cosmologically interesting amounts of cold dark matter.

In this paper we are instead motivated primarily by data rather than by
general theory, mainly by the $ee\gamma\gamma + \slashchar{E}_T$ event
reported by CDF~\cite{park}.  This event has two different possible
supersymmetric interpretations.  One~\cite{dimo96,ambrosanio96} is that
two selectrons are created which ultimately decay into electrons,
photons, and a very light gravitino (less than about $1\kev$) as would happen
in low scale gauge-mediated supersymmetry 
breaking models.  However, these models have not as yet produced a
compelling cold dark matter candidate, and so we do not consider
the light gravitino interpretation further for this paper.

The second supersymmetric interpretation of this
event~\cite{ambrosanio96} is based on ordinary gravity communicated
supersymmetry breaking where the gravitino is sufficiently heavy to not
be the LSP.  In this interpretation, the decay chain which produces
$ee\gamma\gamma + \slashchar{E}_T$ is 
\beq 
\tilde e^+ \tilde e^- \to e^+
N_2 e^- N_2 \to e^+e^-\gamma\gamma \hbeta\hbeta 
\eeq 
where the
photino-like second-lightest neutralino $N_2$ decays
radiatively~\cite{haber89:267} into the lightest neutralino ($\hbeta$)
and a photon.  Thus, $\hbeta$ is the LSP.  Here, motivated by the desire
for a simple notation (and one suitable for e-mail) we denote
neutralinos by $N_i$ (and charginos by $C_i$).  If the supersymmetric
interpretation of the FNAL $ee\gamma\gamma\slashchar{E}_T$ event is
correct, then $\hbeta$ has been observed at FNAL.  Of course, that it
escapes the detector only proves it lives longer than $\sim 10^{-8}$
sec, so its direct detection would be necessary before it could be 
finally accepted
as the cold dark matter.  In this paper we will demonstrate that
$\hbeta$ can be an interesting dark matter candidate, despite the fact
that it is Higgsino-like, and we also discuss the direct detection
prospects for such a particle.  Our analysis assumes a general low scale
supersymmetric Lagrangian, with no presumed relations among parameters.
No assumptions are made about the exact form of the high scale theory
except that it exists perturbatively, and no assumptions are made
about common gaugino or scalar masses.  Rather, we assume that the low
energy theory can be described by a superpotential plus general soft-breaking
terms. We use the results of ref.~\cite{ambrosanio96} for the
mass and coupling requirements of the light supersymmetric states.

\section*{$\hbeta$ as the LSP}

In order to proceed with a discussion about the dark matter qualities
of the $\hbeta$ LSP, we must discuss its composition and mass.
For convention purposes we write down the neutralino
mass matrix
\begin{equation}
\left( 
\begin{array}{cccc}
M_1 & 0 & -M_Z \cos\beta\sin\theta_W & M_Z\sin\beta\sin\theta_W \\
0 & M_2 & M_Z\cos\beta\cos\theta_W & -M_Z\sin\beta\cos\theta_W \\
-M_Z\cos\beta\sin\theta_W & M_Z\cos\beta\cos\theta_W & 0 & -\mu \\
M_Z\sin\beta\sin\theta_W & -M_Z\sin\beta\cos\theta_W & -\mu & 0 
\end{array}
\right)
\end{equation}
in the $\left\{ \tilde B,\tilde W^3,-i\tilde H_d^0,
-i\tilde H^0_u\right\}$ basis.
If $0< -\mu < (M_1\simeq M_2),$ and $\tan\beta$ is near one, then  
the two
lightest eigenstates of the neutralino mass matrix are $N_2 \sim \tilde
\gamma$ (photino-like), and $N_1 \sim
  \sin\beta \tilde H_d^0 + \cos\beta \tilde H_u^0
           +\delta \tilde Z$ (Higgsino-like), where $\delta < 0.1$.
	
This arrangement of lightest neutralino mass eigenstates enhances the
important radiative neutralino decay $N_2 \to \hbeta\gamma$, and along
with the $ee\gamma\gamma+\slashchar{E}_T$ event of ref.~\cite{park}
implies $m_{N_2}-m_{\hbeta}\gsim 30\gev$ and
$30 \lsim m_\hbeta \lsim 55 \gev$~\cite{ambrosanio96}.  We
shall see that with $\tan\beta$ near one the $Z$ invisible width constraint
is satisfied and $\hbeta$ provides an interesting amount of dark matter
as well.

\begin{figure}
\centering
\epsfysize=3.25in
\hspace*{0in}
\epsffile{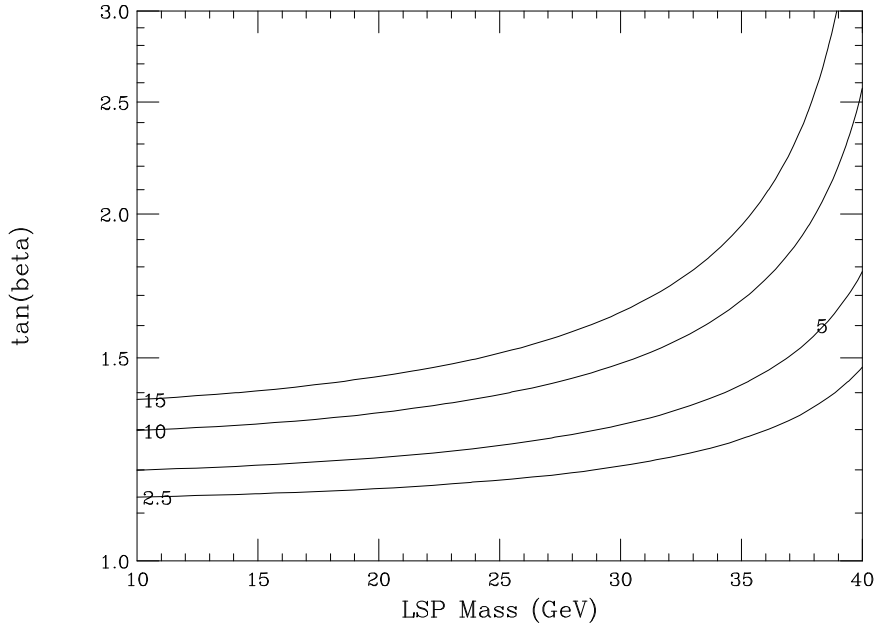}
\caption{Contours of constant invisible width due to $Z\to N_1 N_1$.
The labelled lines are in units of MeV, and the current $2\sigma$
bound at LEP on the invisible width is $5\mev$.}
\label{inv:fig}
\end{figure}

From the invisible width determinations at LEP, the $Z$ is not allowed
to decay into $\hbeta\hbeta$ with a partial width more than about 5 MeV
(at $2\sigma$)~\cite{LEP}.  In our approximation, the formula for the partial 
width is
\begin{equation}
\Gamma_{\mbox{inv}}=\frac{\alpha M_Z}{24 \sin^2\theta_W\cos^2\theta_W}
   \cos^2 2\beta \left( 1-4\frac{m_\hbeta^2}{M_Z^2}\right)^{3/2}.
\end{equation}
Figure~\ref{inv:fig} has contours of the invisible width in units of MeV.
As can be seen from the above equation as long as $\tan\beta$ is
close enough to 1 then the invisible width constraint can be satisfied.
We think this constraint should not be applied too tightly at this stage
since it could be affected by new physics.  However, since $\tan\beta \simeq 1$
is the natural region for the radiative decay requirements
in ref.~\cite{ambrosanio96}, this constraint does not cause
any problems for our analysis here, even if it is applied with the
most stringent assumptions~\cite{feng95}.

$\hbeta$ pairs annihilate through the $Z$ into fermion pairs.  To
look at the prediction for $\Omega h^2$, we can expand the thermally
averaged annihilation cross section~\cite{tavg} into fermions $(f)$ in
the following way:
\begin{equation}
(\sigma v)(x) =\cos^2 2\beta \sum_f (a_f+b_f x)
\end{equation}
where $a_f$ and $b_f$ depend only on one unknown, the LSP mass.
Applying the usual approximation
method~\cite{ellis84:453} to solve the Boltzmann equation, the relic
abundance can be found:
\begin{equation}
\Omega h^2=(2.5\times 10^{-11})\left( \frac{T_\hbeta}{T_\gamma}\right)^3
  \left( \frac{T_\gamma}{2.7\,\mbox{K}}\right)^3 
  \frac{\sqrt{N_F}}{\cos^22\beta}
  \left( \frac{\mbox{GeV}^{-2}}{ax_f+\frac{1}{2}bx_f^2}\right)
\end{equation}
where $N_F$, $(T_\hbeta/T_\gamma)^3$ and $x_f$ must be solved for
self-consistently.  
Calculations such as these could be valid to a
factor of two or better.

\begin{figure}
\centering
\epsfysize=3.25in
\hspace*{0in}
\epsffile{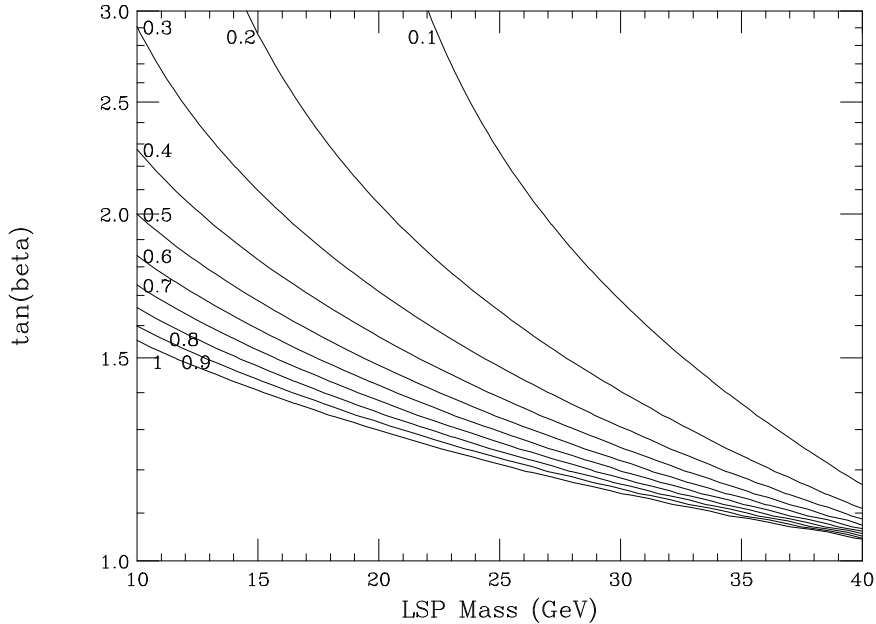}
\caption{Contours of constant $\Omega h^2$ for the Higgsino-like LSP
described in the text.}
\label{omega:fig}
\end{figure}

In Fig.~\ref{omega:fig} contours of $\Omega h^2$ are plotted in the
$\tan\beta - m_\hbeta$ plane.
Since the annihilation cross section is proportional to $\cos^2 2\beta,$
when $\tan\beta$ gets closer to 1, $\Omega h^2$ begins to exceed one.
The $t$ channel sfermion exchange is greatly suppressed (since $\hbeta$
is mainly Higgsino-like), but if the $\tilde Z$ fraction of $\hbeta$
is large enough, then a $t$ channel sfermion diagram which couples like
the $SU(2)_L$ gauge coupling could start to become important.  However,
this potentially efficient annihilation channel is suppressed by a
factor of $\delta^4$ and $\delta$ is less than about
$0.1$~\cite{ambrosanio96}.  Comparing $\delta^4/m^4_{\tilde f}$ with
$\cos^2 2\beta/M_Z^4$, we find that this channel can compete with the
$s$ channel $Z$ exchange only when $\tan\beta$ is less than about
$1.05$.  Furthermore, as $\tan\beta$ gets closer and closer to 1,
$\delta$ necessarily becomes smaller and smaller, and the number
$\tan\beta <1.05$ is actually an overestimate.  At $\tan\beta =1.05$ we
find that the annihilation cross section is too small for all values of
$m_\hbeta$ in Fig.~\ref{omega:fig}, and therefore we set this as our
lower bound on allowed $\tan\beta$ in this scenario.  Consequently, the
sfermion annihilation channel is not numerically important.

The $s$ channel Higgs exchange diagrams might possibly play an important
role in the annihilation cross section.  However, if the pseudo-scalar
($A^0$) is sufficiently heavy then the pseudo-scalar and heavy scalar
($H^0$) Higgses will decouple; in the absence of information about the
heavy Higgs bosons we assume they effectively decouple.  In
this limit it can be shown that the $\hbeta\hbeta h^0$ also vertex decouples.
Furthermore, since the LSP is sufficiently light not to annihilate
into top quarks, or vector bosons, the light final state fermion masses
contribute a further suppression of the $h^0$ mediated annihilation
cross section.  

The allowed mass range for $\hbeta$ from ref.~\cite{ambrosanio96}
overlaps $M_Z/2.$ If consistency with the supersymmetry interpretation
of the LEP $Z\rightarrow b\bar b$ excess is required, probably $M_\hbeta
\lsim 40 \gev$, but it is premature to assume that.  If $M_\hbeta \simeq
M_Z/2$ it is necessary to do the resonant calculation very
carefully~\cite{morekane}.  There is always a value of $\tan\beta$ for
which the curves of Fig. 2 continue across $M_Z/2$ smoothly, so we will
wait until $m_\hbeta$ and $\tan\beta$ are better measured to do the more
precise calculations needed.  We show results in Fig. 2 for $M_\hbeta <
M_Z/2,$ which we expect is the most relevant region.  Given the results
of reference~\cite{ambrosanio96}, 
the only channel that could complicate the simple
analysis is coannihilation of the $\hbeta$ with the 
$\tilde t_1$~\cite{fukugita94}, if
$m_{\tilde t_1} \simeq m_\hbeta$ ($\tilde t_1$ is the lightest stop mass
eigenstate).  We expect $m_{\tilde t_1} \gsim M_Z/2$ and $M_\hbeta \lsim
M_Z/2$, so probably this complication can be ignored, but until the
masses are better determined it should be kept in mind.

The Hubble constant $h$ is probably between about $0.5$ and $0.8$.
Assuming the cold dark matter constitutes 0.4 to 0.8 of
$\Omega_{\hbox{tot}}$, we expect that $\Omega_\hbeta h^2$ should lie
somewhere between $0.08$ and $0.5$ in Fig.~\ref{omega:fig} (e.g. $0.57^2
\times .75 = 0.25$). We emphasize that Fig. 2 follows from the results
of ref.~\cite{ambrosanio96}, and that apart from the approximations
mentioned above this is a prediction of the supersymmetric
interpretation of the CDF event.  Further, we note that the
supersymmetric interpretation of the reported excess of $Z\rightarrow
b\bar b$ decays at LEP leads to the same region of parameters as
ref.~\cite{ambrosanio96}, with Higgsino-like $\hbeta$ and with
$\tan\beta$ near 1~\cite{WK}, and therefore can conservatively be viewed
as consistent with this prediction, or optimistically as additional
evidence for its correctness.

\section*{Detecting $\hbeta$ as the Cold Dark Matter}

We have established that the same $\hbeta$ which is necessary to explain
the $ee\gamma\gamma + \slashchar{E}_T$ event at Fermilab, and
independently the LEP $R_b$ excess, is also a viable cold dark matter
candidate.  Future experiments at Fermilab and LEP will be able to
determine if the supersymmetric interpretation of the CDF event is a
valid one.  However, these colliders cannot determine experimentally if
$\hbeta$ particles in fact are stable and comprise a significant portion
of the cold dark matter in the universe.  In this section we discuss
some of the direct detection prospects for this particle.

From kinematic analyses of 
the galactic rotation curves it has been
estimated~\cite{gates95:039} that the local density of cold dark matter
is approximately $0.3 < \rho < 0.7\gev/\hbox{cm}^3$.  (We will use the
lower number in our subsequent calculations.) 
Several experiments are under way to look for weakly interacting massive
particles (WIMPs) floating around our part of the galaxy.  Neutrino
telescopes at AMANDA, etc., hope to see the effects of WIMP
annihilations in the sun.  The light Higgsino dark matter candidate,
which we propose here, would be difficult to detect at the large area
neutrino telescopes since the muon threshold energy is about $30\gev$,
roughly equivalent to the LSP mass range we are considering.  The neutrinos
produced by the LSP annihilations in the sun will be at energies 
below this threshold, and so the converted muons will not be energetic
enough to be detected.

Other experiments~\cite{tarle} are designed
to measure direct annihilations of LSPs in the galactic halo by seeing
an excess of photons, electrons, protons, etc.\ in the spectrum.  
The steeply rising photon background (as energy decreases) makes the
photon signal difficult to extract, and the broad energy spread of the
electron/positron signal for $N_1N_1$ annihilations in the galactic
halo also complicates this detection possibility.

Lastly, numerous table-top experiments~\cite{lsptt} 
are being set up with the hope of
seeing WIMPs interact with different nuclei.  It is this last set of
experiments that we wish to focus on in this analysis.  We also wish to
urge one change in notation, namely that the acronym WISP (weakly
interacting supersymmetric particle) be used when the particle in
question is known to be a possible state following from a supersymmetric
Lagrangian and consistent with phenomenological constraints; many WIMP's
discussed in the literature are not WISP's.

Since we have argued that the $Z$ coupling is the most important one, we
concentrate on ${}^{29}\mbox{Si}$ and ${}^{73}\mbox{Ge}$ 
which are well-suited for spin-dependent
scattering of LSP's with nucleons, and are currently being considered by
experimentalists for larger scale designs.  The spin-dependent cross
section in this scenario can be written as
\begin{equation}
\sigma^{\mbox{sd}}=\frac{g_2^4}{16\pi}\frac{\cos^2 2\beta}{M_W^4}
      \frac{m_\hbeta^2m_A^2}{(m_\hbeta+m_A)^2}\lambda^2 J(J+1)
      [\Delta_n u -\Delta_n d-\Delta_n s]^2
\end{equation}
where $\Delta_n q$ is the spin content of the neutron~\cite{ellis93:131} 
carried by quark $q$,
and $m_A$ is the mass of a silicon atom.
From this we can estimate~\cite{lsptt} 
the rate of interactions per day:
\begin{equation}
R=\frac{\sigma \xi}{m_\hbeta m_A}\left( 
 \frac{1.8\times 10^{11}\mbox{GeV}^4}{\mbox{kg}\cdot \mbox{day}}\right)
\end{equation}
where $\xi$ quantifies the nuclear form factor suppression.
We are not considering the spin-independent cross section in this
analysis since, as we argued above, the couplings of $N_1$ to all
scalar particles are very small.  It is possible with a lighter 
pseudo-scalar mass to
have larger couplings of $N_1$ to the Higgs particles, which would 
contribute to a spin-independent cross-section, but to be conservative
we have assumed that the Higgs effects are decoupled.

\begin{figure}
\centering
\epsfysize=3.25in
\hspace*{0in}
\epsffile{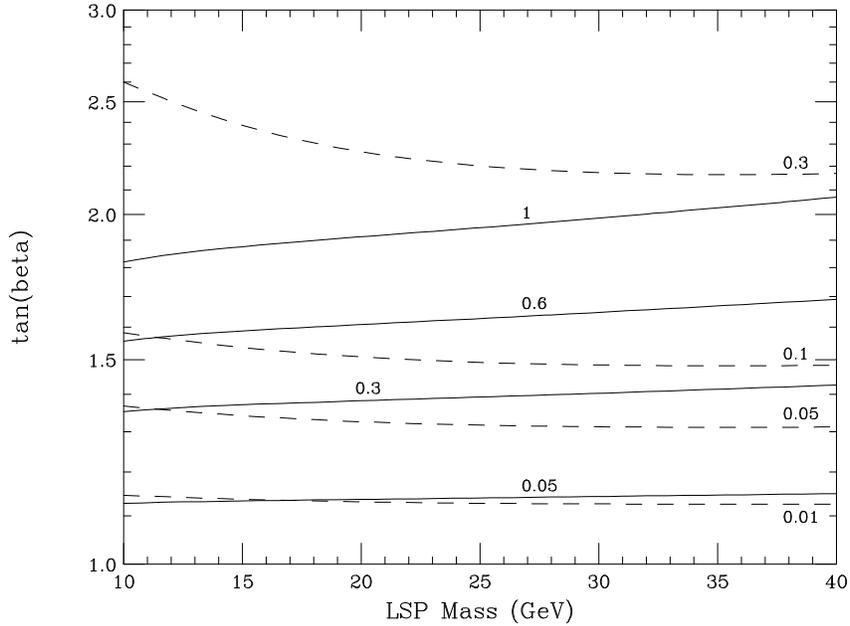}
\caption{Event rate contours for ${}^{29}\mbox{Si}$ (solid lines)
and ${}^{73}\mbox{Ge}$ (dashed lines) in units of /kg/day.}
\label{si29:fig}
\end{figure}

Figure~\ref{si29:fig} is a plot of the event rate per kilogram per day
of $\hbeta$ interacting on ${}^{29}\mbox{Si}$ and ${}^{73}\mbox{Ge}$.
The expected sensitivity~\cite{ge73} 
for ${}^{73}\mbox{Ge}$ is at about the $0.3$
events contour in the near future, and about $0.01$ in the next round
of experiments.  Thus, the entire region of the plot above about 
the $0.01$ event contour will soon be probed in the table-top detector.  
This is a
good demonstration of how the table top experiment can sometimes do
better than collider limits (see Fig.~\ref{inv:fig}) for an interesting
part of parameter space.  
${}^{73}$Ge has a reduced event rate
compared to Silicon mainly because the nuclear Land\'e $\lambda^2
J(J+1)$ factor is smaller, and the nucleus mass is heavier.  
We interpret Fig.~\ref{si29:fig} as implying that Si and
Ge and related detectors may be able to observe a cold dark matter
signal in the next round of attempts.

\section*{Summary}

It is remarkable that these calculations (which \hbox{\it a priori}\/ 
could have given much
smaller or much larger $\Omega h^2,$ and are essentially free of
parameters) imply a WISP cold dark matter candidate that is
cosmologically interesting, possibly detectable using table-top
experiments, possibly already observed at FNAL, and possibly associated
with loop effects seen at LEP.

\section*{Acknowledgements}

We would like to thank S.~Ambrosanio, B.~Cabrera, D.~Caldwell, E.~Diehl,
G.~Kribs, S.~Martin and M.~Worah 
for useful discussions.  This work was supported 
in part by the U.S. Department of Energy.

\vfill\eject

\end{document}